\documentclass[11pt]{article}
\usepackage{amsmath,amssymb,color,graphics,epsfig,cite}

\usepackage{amsfonts}

\newcommand{\hoch}[1]{$\, ^{#1}$}

\makeatletter
\@addtoreset{equation}{section}
\makeatother

\begin{document}

\begin{flushright} 

\hfill{}

\end{flushright}

\vspace{25pt}
\begin{center}
{\Large {\bf Symmetries at the black hole horizon}}

\vspace{15pt}
{\bf Emil T. Akhmedov\hoch{1} and Mahdi Godazgar\hoch{2}}

\vspace{10pt}

\hoch{1} {\it Institutskii per, 9, Moscow Institute of Physics and Technology, 141700, Dolgoprudny, Russia \\
B. Cheremushkinskaya, 25, Institute for Theoretical and Experimental Physics, 117218, Moscow, Russia
 \\
akhmedov@itep.ru}

\vspace{10pt}

\hoch{2} {\it Institut f\"ur Theoretische Physik,\\
Eidgen\"ossische Technische Hochschule Z\"urich, \\
Wolfgang-Pauli-Strasse 27, 8093 Z\"urich, Switzerland.\\
godazgar@phys.ethz.ch}

\vspace{20pt}
July 18, 2017

\vspace{20pt}

\underline{ABSTRACT}
\end{center}
We determine the asymptotic symmetry group of Killing horizons by choosing Gaussian null coordinates in the neighbourhood of the horizon and boundary conditions that respect the leading order terms in the metric.  The analysis divides naturally into the two cases of subextremal and extremal horizons. In general, we find rather involved asymptotic symmetry generators that nevertheless involve supertranslations and Killing vectors on the compact horizon.  The most striking observation is the difference in the dependence on the coordinate along the horizon; we relate this to the redshift effect for subextremal horizons.  We consider the spherically symmetric case as a special and illuminating example.


\thispagestyle{empty}

\pagebreak




\section{Introduction}

Identifying special properties and structures of black hole horizons is considered to be an important step in solving long-standing problems in black hole physics. The appearance of Virasoro symmetries in the near-horizon geometry of extreme black holes has lead to arguments along the lines of Brown and Henneaux \cite{Brown:1986nw} regarding the countability of black hole microstates and the relation to conformal field theory states (see e.g.\ Refs.\ \cite{Carlip:1999cy, Guica:2008mu, Seraj:2016cym}).  

Following Hawking, Perry and Strominger \cite{HPS, HPS2}, much recent attention has focused on the black hole information paradox and its possible clarification using asymptotic symmetries (see Refs.\cite{HPS, HPS2} and references therein).  Following Refs. \cite{HPS, HPS2}, the majority of the literature on this subject has proceeded its investigations by either assuming or arguing for the existence of the full BMS symmetry group on the horizon of a black hole, or at least its Abelian subgroup comprising supertranslations \cite{Donnay:2015abr, Blau:2015nee, Averin:2016ybl, Sheikh:2016lzm, Mirbabayi:2016axw, Cai:2016idg, Shi:2016jtn, Carlip:2017xne, Hou:2017pes}. 

In this paper, we determine the asymptotic symmetry group at the horizon of an arbitrary four-dimensional black hole~\footnote{As opposed to the case of asymptotic flatness, where the BMS group does not exist in higher dimensions \cite{Tanabe:2011es}, for the horizon, very similar results can be derived in higher dimensions.  However, for clarity, here, we focus our attention on four-dimensional black holes.}, much in the same way as the BMS symmetry group is generally derived for asymptotically flat spaces: for any asymptotically flat space, one can choose Bondi coordinates $(u,r,x^I)$ in which the metric takes a Bondi-Sachs form \cite{bondi, sachs}.  The Bondi coordinates are so chosen such that for large values of the null radial coordinate $r$, the metric approaches flat space.~\footnote{This rather vague-seeming statement can be made mathematically precise by appealing to Penrose's conformal compactification \cite{Geroch1977, Ashtekar:1987tt}.}  The asymptotic symmetry group, the BMS group, is then found by finding asymptotic isometry generators such that the asymptotic form of the metric remains preserved, i.e. the metric remains asymptotically flat.

Whereas for asymptotically flat solutions the background about which one is to perform an asymptotic symmetry analysis of the type described above is clear, this is not so for back hole spaces as one approaches their horizon.  The geometry near a black hole horizon is not, in general, Ricci-flat.  Nor is there a peeling property (see e.g.\ \cite{wald}) of the sort that exists as one approaches asymptotic null infinity for black hole horizons.  One option would be to perform such an analysis for each black hole horizon individually, say for the horizon of a Schwarzschild black hole, with the background metric chosen as the Schwarzschild metric to leading order at the horizon \cite{Hotta:2000gx, Koga:2001vq}.  However, it would be more satisfactory to be able to make a general statement about any black hole horizon. 

Assuming the horizon to be Killing, one can always introduce Gaussian null coordinates $(v,r,x^I)$ in the neighbourhood of the horizon \cite{Friedrich:1998wq, Kunduri:GNC}, which is the null surface at $\{r=0\}$, such that the metric takes a universal form depending on whether the horizon is bifurcate ($n=1$ in the metric below) (subextremal) or not ($n=2$ in the metric below)
\begin{equation}
 ds^2 = L(x)^2 \Big[- r^n F(r,x) dv^2 + 2 dv dr \Big]  + \gamma_{IJ}(r,x) \Big(dx^I - r h^I(r,x) dv \Big) \Big(dx^J - r h^J(r,x) dv\Big).
\end{equation}
We take this metric to define the background space on which we perform our asymptotic symmetry analysis.  Moreover, we choose boundary conditions for the metric components that preserve the leading order components in the metric. Thus, any black hole that lies within a class in which the leading order components of the metric are the same, has a particular asymptotic symmetry group at the horizon shared by all black holes in the same class.  This contrasts, for example, with the analysis of Ref.\ \cite{Eling:2016xlx}, where they assume weaker boundary conditions and, therefore, obtain a larger symmetry group.  In general, for both extremal and subextremal horizons, we find supertranslations at the horizon, in agreement with previous such analyses \cite{Hotta:2000gx, Koga:2001vq, Koga:2006ez, Koga:2006sb, Eling:2016xlx, Donnay:2016ejv, Hou:2017pes}. 

We consider the two cases of extremal and subextremal horizons separately.  For extremal horizons we find that the asymptotic symmetry generator is of the form
\begin{gather}
 \xi = \left[ f(x) + g(x) v + h(x) v^2 \right] \partial_v + \xi^I \partial_I \hspace{70mm} \notag \\[1mm]
\hspace{16mm} + \left\{ \hat{\xi}^r(x) - (g + 2 h\,  v) \, r - \int dr \left[ 2\, \xi \cdot \partial \log L + r h^I \partial_I (f+gv+hv^2) \right]  \right\} \partial_r,
\end{gather}
where 
\begin{equation}
 \xi^I = \alpha^I(x) + \beta^I(x) v - L^2 \partial_{J} (f+gv+hv^2) \int dr\, \gamma^{IJ}.
\end{equation}
Hence, we have four $x$-dependent functions $f, g, h$ and $\hat{\xi}^r$ and two vectors $\alpha^I$ and Killing vector $\beta^I$ on the compact space given by coordinates $x^I$.  The Killing vector $\beta^I$ and $h$ are given in terms of $\hat{\xi}^r$, see equations \eqref{2:beta} and \eqref{2:h}, which is itself constrained by two linear equations, \eqref{2:xir1} and \eqref{2:xir2}.  Moreover, $\hat{\xi}^r$, $g$ and $\alpha^I$ solve two further linear equations, \eqref{2:xiga1} and \eqref{2:xiga2}.  Importantly, function $f$ is arbitrary and is most identifiable as a supertranslation on the horizon.  In summary, we have at most three independent function $f, g$ and $\hat{\xi}^r$, which generates Killing vectors $\beta^I$, and a vector $\alpha^I$.

On the other hand, for a subextremal horizon, with non-zero surface gravity $\kappa$, the asymptotic symmetry generator is of the form
\begin{gather}
 \xi = \big[ f(x) e^{- \kappa v} + g(x) + h(x) e^{\kappa v} \big] \partial_v + \xi^I \partial_I \hspace{70mm} \notag \\[1mm]
\hspace{10mm} + \left\{ \zeta(x)\, e^{\kappa v} +\frac{1}{2} \left( f(x) e^{- \kappa v} - h(x) e^{\kappa v} \right) \, r - \int dr \left[ 2\, \xi \cdot \partial \log L + r h^I \partial_I \xi^v \right] \right\} \partial_r,
\end{gather}
where 
\begin{equation}
 \xi^I = Y^I(x) + 2\,  e^{\kappa v} \left( \tilde{h}^I \zeta - L^2 \omega^{IJ} \partial_{J} \zeta \right) - L^2 \partial_J \xi^v \int dr\, \gamma^{IJ}
\end{equation}
with $\tilde{h}^I(x)$ and $\omega_{IJ}$ the leading order terms in the $r$ expansions of $h^I$ and $\gamma_{IJ}$, respectively.  This time we have four $x$-dependent functions $f, g, h$ and $\zeta$ and a Killing vector $Y^I.$  Function $h$ is given in terms of $\zeta,$ equation \eqref{1:h}, which is in turn constrained by two linear equations, \eqref{1:zeta1} and \eqref{1:zeta2}.  Moreover, the Killing vector $Y^I$ and $g$ satisfy a linear equation, \eqref{1:gY}. As before, $f(x)$ is an arbitrary function and can most easily be identified with supertranslations on the horizon.

The most striking difference with the extremal case is the difference in the $v$-dependence of the generators.  Whereas, for the extremal case the asymptotic symmetry generators have a polynomial dependence on $v$, for the subextremal case the dependence is exponential.  This is a novel feature of the asymptotic symmetry generators found here compared to those proposed in the literature before.
  
The above expressions give the asymptotic symmetry generators for the most general horizons.  However, as mentioned earlier, due to the boundary conditions chosen, we can define an asymptotic symmetry group for each class of horizons sharing the same leading order metric components in Gaussian null coordinates.  Specialising to spherically symmetric backgrounds, as well as assuming that $\tilde{h}^I$, the leading order term in the $r$ expansion of $h^I$, vanishes, gives us a class of horizons that includes the Reissner-Nordstr\"om (and Schwarzschild) family of black holes.  In this case, apart from the distinct $v$-dependence discussed earlier, the asymptotic symmetry groups for both the extremal and subextremal horizons are generated by a constant $c$, a supertranslation $f(x)$ and the three Killing vectors $Y^I$ on the round 2-sphere.

In section \ref{sec:ASG}, we derive the asymptotic symmetry generators for a general horizon, treating the two cases of subextremal and extremal horizons separately.  In section \ref{sec:kappa}, we give a partial physical explanation of the strikingly different $v$-dependence in the two cases by observing that the $v$-dependence of the asymptotic symmetry generator for the subextremal case is the same as the well-known redshift factor that one encounters when approaching a bifurcate horizon.  There is no such effect for extremal horizons.  In section \ref{sec:sph}, we consider as an example the case of spherically symmetry, leading to a very simple statement for the asymptotic symmetry group.  Finally, we end with some discussions in section \ref{sec:dis}.

\section{Asymptotic symmetries at the horizon} \label{sec:ASG}

Before we consider the horizon, let us recall briefly how one finds the asymptotic symmetry group of asymptotically flat spaces.  For any asymptotically flat space, we may introduce Bondi coordinates $(u,r,x^I=\{\theta,\phi\})$ \cite{bondi, sachs}, such that the metric takes the form
\begin{equation} \label{AF}
 d s^2 = - F e^{2 \beta} du^2 - 2 e^{2 \beta} du dr + r^2 h_{IJ} (dx^I - C^I du) (dx^J - C^J du),
\end{equation}
The metric functions satisfy the following fall-off conditions for large $r$:
\begin{gather}
 \lim_{r\to\infty} r(F-1) = \tilde{F}(u,x), \qquad \lim_{r\to\infty} r^2 \beta = \tilde{\beta}(u,x), \notag \\[2mm] \label{falloff}
 \lim_{r\to\infty} r^2 C^I = \tilde{C}^I(u,x), \qquad \lim_{r\to\infty} r(h_{IJ}-\omega_{IJ}) = \tilde{h}_{IJ}(u,x),
\end{gather}
where the 2-dimensional space given by metric $\omega_{IJ}$ is the standard metric on the round unit sphere.  Moreover, we can always choose a gauge in which
\begin{equation} \label{det:h}
 h \equiv \textup{det}(h_{IJ}) = \omega = \sin\theta.
\end{equation}

Now, we find the asymptotic symmetry by imposing that the variation of the metric under the generators of the asymptotic symmetry group respect the form of the metric and the gauge choices.
These conditions imply that
\begin{equation}
 \xi = f \partial_u +  \left( Y^I + \int dr \frac{e^{2\beta}}{r^2} h^{IJ} \nabla_{J} f \right) \partial_I - \frac{r}{2} \left( \nabla_I \xi^I - C^I \nabla_I f \right) \partial_r,
\end{equation}
where $\nabla_I$ is a covariant derivative associated with metric $\omega_{IJ}$.  The $u$ and $r$-independent function $f(z, \bar{z})$ is called a supertranslation, while the conformal Killing vectors on the 2-sphere, $Y^I$, generate the compact part of the BMS group.  

Now, let us consider the analogous situation for the horizon.  First, we need to define an asymptotic form of the metric in this setting.  Given a stationary black hole metric with a Killing horizon, we may introduce coordinates $(v,r,x^I)$ such that the horizon is at $r=0$ and the metric takes the form \cite{Friedrich:1998wq, Kunduri:GNC}
\begin{equation} \label{NHC}
 ds^2 = L(x)^2 \Big[- r^n F(r,x) dv^2 + 2 dv dr \Big]  + \gamma_{IJ}(r,x) \Big(dx^I - r h^I(r,x) dv \Big) \Big(dx^J - r h^J(r,x) dv\Big)
\end{equation}
with $F(r=0,x) = 1$ and $\gamma_{IJ}(0,x) = \omega_{IJ},$ where, here, $\omega_{IJ}$ is the metric corresponding to some compact space.  Moreover, we have that 
\begin{equation} \label{h2falloff}
  \lim_{r \to 0} h^{I} = \tilde{h}^I(x).
\end{equation}
In the $vv$ component above, the integer power
\begin{equation}
 n = \begin{cases}
      1 & \textup{subextremal horizon} \\
      2 & \textup{extremal horizon}
     \end{cases}.
\end{equation}

We repeat the asymptotic symmetry analysis as before.  That is, we look for diffeomorphism generators $\xi$, under which the form of the metric is preserved.  Under a diffeomorphism generator $\xi$, the variation of the metric
\begin{equation}
 \delta_{\xi} g_{ab} = (\mathcal{L}_{\xi} g)_{ab} = \xi^c \partial_{c} g_{ab} + g_{ca} \partial_{b} \xi^c + g_{cb} \partial_{a} \xi^c.
\end{equation}
Assuming that the metric functions admit a Taylor expansion, at least for the first couple of terms,
\begin{gather}
F = 1 + r\, F^{(1)}(x) + \mathcal{O}(r^2), \quad h^{I} = \tilde{h}^I(x) + r\, h^{(1)I}(x)  + \mathcal{O}(r^2), \notag \\[1mm]
\gamma_{IJ} = \omega_{IJ} + r\, \gamma^{(1)}_{IJ}(x)  + \mathcal{O}(r^2),
\end{gather}
we impose that
\begin{gather}
 \delta_{\xi} g_{rr} = 0, \qquad \delta_{\xi} g_{rI} = 0, \qquad \delta_{\xi} g_{rv} = 0, \notag \\ \label{met:bdry}
 \delta_{\xi} g_{vv} = \mathcal{O}(r^{n+1}), \qquad \delta_{\xi} g_{vI} = \mathcal{O}(r^2), \qquad \delta_{\xi} g_{IJ} = \mathcal{O}(r).
\end{gather}
The above boundary conditions that we impose on the variation of the metric under the asymptotic symmetry generator ensure that $L(x),$ $\tilde{h}^{I}$ and $\omega_{IJ}$ remain unchanged under the transformation.  For the extremal case ($n=2$), this would imply that the near-horizon geometry \cite{Kunduri:GNC} remains invariant under the action of the asymptotic symmetry group, as one would expect.

It is simple to find that the conditions on the first line of \eqref{met:bdry} imply that the asymptotic symmetry generators must be of the form
\begin{gather}
 \xi^v = \xi^v(v,x), \label{xi:v} \\[2mm]
 \xi^I = \hat{\xi}^{I}(v,x) - L^2 \partial_J \xi^v \int dr\, \gamma^{IJ}, \label{xi:I} \\[2mm]
 \xi^r = \hat{\xi}^{r}(v,x) - \partial_v \xi^v \, r - \int dr \left[ 2\, \xi \cdot \partial \log L + r h^I \partial_I \xi^v \right], \label{xi:r}
\end{gather}
where $\gamma^{IJ}$ is the inverse of $\gamma_{IJ}$
\begin{equation}
 \gamma^{IK} \gamma_{KJ} = \delta^I_J
\end{equation}
and, for brevity, we use the notation $\xi \cdot \partial \equiv \xi^I \partial_I$.
Expanding equations \eqref{xi:I} and \eqref{xi:r} as power series in $r$ gives
\begin{gather}
 \xi^I = \hat{\xi}^I - r\ L^2 \omega^{IJ} \partial_J \xi^v  + \mathcal{O}(r^2), \label{xi:Ir} \\[2mm]
 \xi^r = \hat{\xi}^r -r\ \left( \partial_v \xi^v + 2\, \hat{\xi} \cdot \partial \log L \right) + \frac{r^2}{2} \left(L^2 \omega^{IJ} \partial_I \xi^v \partial_J \log L - \tilde{h} \cdot \partial \xi^v \right)  + \mathcal{O}(r^3), \label{xi:rr}
\end{gather}
where $\omega^{IJ}$ is the inverse of $\omega_{IJ}$ and, therefore, $r$-independent.

In order to make progress with the other conditions in \eqref{met:bdry}, we need to distinguish between $n=1$ and $n=2$.  We start by analysing the $n=2$ case, which corresponds to the geometry in the vicinity of an extremal horizon.

\subsection*{n=2: Extremal case}

Expanding $\delta_\xi g_{vv} = \mathcal{O}(r^{3})$ ($Eq_{vv}$) in powers of $r$, the only $r$-independent term is
\begin{equation*}
\partial_v \hat{\xi}^r.
\end{equation*}
Thus,
\begin{equation} \label{2:xirdef}
 \hat{\xi}^r = \hat{\xi}^{r}(x).
\end{equation}
Using the above result, the $r$-independent terms in $\delta_\xi g_{vI} = \mathcal{O}(r^{2})$ ($Eq_{vI}$) imply that
\begin{equation} \label{2:Idef}
 \hat{\xi}^I = \alpha^I(x) + \beta^I(x) \, v,
\end{equation}
where
\begin{equation} \label{2:beta}
 \beta^I = \tilde{h}^I \hat{\xi}^r - L^2 \omega^{IJ} \partial_{J} \hat{\xi}^r.
\end{equation}
Moreover, constraint $\delta_{\xi} g_{IJ} = \mathcal{O}(r)$ ($Eq_{IJ}$) implies that $\beta^I$ is a Killing vector field on the background defined by the metric $\omega_{IJ}$
\begin{equation} \label{2:betaKilling}
 \mathcal{L}_\beta \omega_{IJ} = 0
\end{equation}
and that
\begin{equation} \label{2:aKilling}
  \mathcal{L}_\alpha \omega_{IJ} + \hat{\xi}^r \gamma^{(1)}_{IJ} = 0.
\end{equation}

Next, using the above results, the linear in $r$ terms in $Eq_{vv}$ give
\begin{equation} \label{2:vdef}
 \xi^v = f(x) + g(x) v + h(x) v^2,
\end{equation}
where
\begin{equation} \label{2:h}
 h = - \frac{1}{2} \left\{ \hat{\xi}^r + \tilde{h} \cdot \partial \hat{\xi}^r + 2 \hat{\xi}^r \tilde{h} \cdot \partial \log L 
 - L^{-2} \hat{\xi}^r \omega_{IJ} \tilde{h}^I \tilde{h}^J - 2 L^2 \omega^{IJ} \partial_{I} \log L \partial_J \hat{\xi}^r \right\},
\end{equation}
Note that as with $\beta^I$, $h$ depends linearly on $\hat{\xi}^r$. 

Finally, the quadratic in $r$ terms in $Eq_{vv}$ and the linear terms in $Eq_{vI}$ give two equations each, which further constrain $\hat{\xi}^r$, $\alpha^I$ and $g$.  These equations are not very illuminating.  However, for completeness, we write them down here.  $Eq_{vv}$ gives
\begin{gather}
 \beta \cdot \partial \log L + \tilde{h} \cdot \partial h + 2 L^2 \omega^{IJ} \partial_I \log L \partial_J h = 0, \label{2:xir1} \\[2mm]
 3 L^2 F^{(1)} \hat{\xi}^r + 2 \gamma^{(1)}_{IJ} \tilde{h}^I \beta^J + 2 \omega_{IJ} h^{(1)I} \beta^{J} - L^2 \tilde{h} \cdot \partial g 
 - 2 L^4 \omega^{IJ} \partial_{I} \log L \partial_J g - 2 L^2 \alpha \cdot \partial \log L = 0, \label{2:xiga1}
\end{gather}
while $Eq_{vI}$ gives
\begin{gather}
 4 L^2 \partial_I h + \beta \cdot \partial (\omega_{IJ} \tilde{h}^J) + 2 L^2 \partial_{I} (\beta \cdot \partial \log L) + \omega_{JK} \tilde{h}^J \partial_I \beta^K 
 - 2 \omega_{IJ} \tilde{h}^J \beta \cdot \partial \log L = 0,  \label{2:xir2} \\[2mm]
 2 \hat{\xi}^r \left(\gamma^{(1)}_{IJ} \tilde{h}^J + \omega_{IJ} h^{(1)J} \right) - \gamma^{(1)}_{IJ} \beta^J + 2 L^2 \partial_I g \hspace{70mm} \notag \\[1mm] 
 \hspace{10mm} + \alpha \cdot \partial (\omega_{IJ} \tilde{h}^J) + 2 L^2 \partial_{I} (\alpha \cdot \partial \log L)  + \omega_{JK} \tilde{h}^J \partial_I \alpha^K 
 - 2 \omega_{IJ} \tilde{h}^J \alpha \cdot \partial \log L=0.
 \label{2:xiga2}
\end{gather}

In summary, we have
\begin{gather}
 \xi = \left[ f(x) + g(x) v + h(x) v^2 \right] \partial_v + \xi^I \partial_I \hspace{70mm} \notag \\[1mm]
\hspace{16mm} + \left\{ \hat{\xi}^r(x) - (g + 2 h v) \, r - \int dr \left[ 2\, \xi \cdot \partial \log L + r h^I \partial_I (f+gv+hv^2) \right]  \right\} \partial_r,
\end{gather}
where 
\begin{equation}
 \xi^I = \alpha^I(x) + \beta^I(x) v - L^2 \partial_{J} (f+gv+hv^2) \int dr\, \gamma^{IJ}.
\end{equation}
The $x$-dependent function $f$ is arbitrary, while $\beta^I$ and $h$ depend linearly on $\hat{\xi}^r$ (see equations \eqref{2:beta} and \eqref{2:h}). $\hat{\xi}^r$ itself is constrained by the fact that $\beta^I$ is a Killing vector field \eqref{2:betaKilling} and equations \eqref{2:xir1} and \eqref{2:xir2}.  Moreover, $\hat{\xi}^r$, $g$ and $\alpha^I$ satisfy equations \eqref{2:xiga1} and \eqref{2:xiga2}.

In order to make the expressions more clear, we can consistently set the constrained functions
\begin{equation}
 \hat{\xi}^r = g = \alpha^I = 0.
\end{equation}
This leaves us with one unconstrained function $f(x)$ and the expression for $\xi$ simplifies to give
\begin{equation}
 \xi = f\, \partial_v + \xi^I \partial_I -  \int dr \left[ 2\, \xi \cdot \partial \log L + r h^I \partial_I f \right]  \, \partial_r
\end{equation}
with
\begin{equation}
 \xi^I =  - L^2\, \partial_{J} f \int dr\, \gamma^{IJ}. 
\end{equation}
We may then identify $f(x)$ as a \emph{supertranslation} generator at the horizon.

\subsection*{n=1: Subextremal case}

In this case, crucially, we have two $r$-independent terms in $Eq_{vv}$, which give
\begin{equation}
 2 \partial_v \hat{\xi}^r + \hat{\xi}^r = 0.
\end{equation}
Hence,
\begin{equation} \label{1:xirdef}
 \hat{\xi}^r = \zeta(x)\, e^{v/2}.
\end{equation}
Using the above result, the $r$-independent terms in $Eq_{vI}$ imply that
\begin{equation} \label{1:Idef}
 \hat{\xi}^I = Y^I(x) + 2\,  e^{v/2} \left( \tilde{h}^I \zeta - L^2 \omega^{IJ} \partial_{J} \zeta \right).
\end{equation}
$Eq_{IJ}$ then implies that $Y^I$ is a Killing vector field on the compact space with metric $\omega_{IJ}$
\begin{equation} \label{1:Y}
 \mathcal{L}_{Y} \omega_{IJ} = 0
\end{equation}
and that 
\begin{equation} \label{1:zeta1}
 \zeta \, \gamma^{(1)}_{IJ} + 2 \mathcal{L}_{X} \omega_{IJ} = 0,
\end{equation}
where
\begin{equation} \label{1:X}
 X^I =  \tilde{h}^I \zeta - L^2 \omega^{IJ} \partial_{J} \zeta.
\end{equation}

Now, using the above results, the linear terms in $Eq_{vv}$ give that
\begin{equation} \label{1:vdef}
 \xi^v = f(x) e^{-v/2} + g(x) + h(x) e^{v/2},
\end{equation}
where
\begin{equation} \label{1:h}
 h =  - 2 \left\{ F^{(1)} \zeta - \tilde{h} \cdot \partial \zeta + 2\, \zeta\, \tilde{h} \cdot \partial \log L 
 + L^{-2} \zeta \omega_{IJ} \tilde{h}^I \tilde{h}^J - 2 L^2 \omega^{IJ} \partial_{I} \log L \partial_J \zeta \right\}.
\end{equation}
Finally, the linear terms in $Eq_{vI}$ give two further equations
\begin{gather}
  L^2 \partial_I g + Y \cdot \partial (\omega_{IJ} \tilde{h}^J) + 2 L^2 \partial_{I} (Y \cdot \partial \log L) + \omega_{JK} \tilde{h}^J \partial_I Y^K \hspace{50mm}  \notag \\[1mm] 
  \label{1:gY}
 \hspace{60mm} - 2 \omega_{IJ} \tilde{h}^J Y \cdot \partial \log L + 2 L^2 \partial_I (Y \cdot \log L) = 0, \\[2mm]
 3 L^2 \partial_I h + 4 \zeta \left(\gamma^{(1)}_{IJ} \tilde{h}^J + \omega_{IJ} h^{(1)J} \right) - 2 \gamma^{(1)}_{IJ} X^J + 4 X \cdot \partial (\omega_{IJ} \tilde{h}^J) \hspace{50mm} \notag \\[1mm] 
 \hspace{20mm} + 8 L^2 \partial_{I} (X \cdot \partial \log L)  + 4 \omega_{JK} \tilde{h}^J \partial_I X^K 
 - 8 \omega_{IJ} \tilde{h}^J X \cdot \partial \log L = 0.
 \label{1:zeta2}
\end{gather}
Note that the first equation above involves only $g$ and $Y^I$. On the other hand, using equations \eqref{1:X} and \eqref{1:h}, the second equation involves only $\zeta$.

In summary, we have
\begin{gather}
 \xi^v = f(x) e^{-v/2} + g(x) + h(x) e^{v/2}, \label{1:xiv} \\[2mm]
 \xi^I = Y^I(x) + 2\,  e^{v/2} \left( \tilde{h}^I \zeta - L^2 \omega^{IJ} \partial_{J} \zeta \right) - L^2 \partial_J \xi^v \int dr\, \gamma^{IJ}, \\[2mm]
 \xi^r = \zeta(x)\, e^{v/2} +\frac{1}{2} \left( f(x) e^{-v/2} - h(x) e^{v/2} \right) \, r - \int dr \left[ 2\, \xi \cdot \partial \log L + r h^I \partial_I \xi^v \right], 
  \label{1:xir}
\end{gather}
where $f(x)$ is an arbitrary function, $h$ is a function of $\zeta$, as given in equation \eqref{1:h}, which is in turn constrained by equations \eqref{1:zeta1} and \eqref{1:zeta2}. Moreover, $Y^I$ is a Killing vector field and with $g$ satisfies equation \eqref{1:gY}.

As before, the asymptotic symmetry generator simplifies significantly if we choose
\begin{equation}
\zeta = g = Y^I = 0,
\end{equation}
which is consistent with the equations above.  In this case,
\begin{equation}
 \xi = e^{-v/2} \left\{  f\, \partial_v + \xi^I \partial_I  + \left( \frac{1}{2} f -  \int dr \left[ 2\, \xi \cdot \partial \log L + r h^I \partial_I f \right] \right) \, \partial_r \right\}
\end{equation}
with
\begin{equation}
 \xi^I =  - L^2\, \partial_{J} f \int dr\, \gamma^{IJ}. 
\end{equation}
As before, we identify function $f$ as a supertranslation generator at the horizon.

\section{Surface gravity} \label{sec:kappa}

We have found rather novel symmetries at play at the horizon of both extremal and subextremal horizons.  Generically, these involve Killing vectors of the compact space with metric $\omega_{IJ},$ which describes the geometry of the horizon and, importantly, supertranslations.  In this section we deal with the striking $v$-dependence of the asymptotic symmetry generators in the two different cases.  For extremal horizons, $\xi$ has polynomial dependence on the coordinate $v$.  However, for subextremal horizons, the $v$-dependence becomes exponential.  This feature is reminiscent of the redshift at subextremal horizons given by
\begin{equation}
 e^{\pm \kappa v},
\end{equation}
where $\kappa$ is the surface gravity.  Of course, for extremal horizons $\kappa = 0$ and, therefore, an external observer does not experience a redshift effect.  In the following, we try to make this relation a bit more precise.

Recall that the surface gravity $\kappa$ associated with a Killing horizon $\mathcal{H}$ generated by Killing vector $k$ is given by
\begin{equation} \label{kappa}
 d (k^2) |_{\mathcal{H}} = -2 \kappa\, k^{\flat}|_{\mathcal{H}}.
\end{equation}
However, there is an ambiguity in the definition above: if $k$ generates the Killing horizon, then so does $c\, k$ for $c$ a constant.  Since in the equation above the left hand side is quadratic in $k$, while the right hand side is linear, this ambiguity can be used to set $\kappa=1$ for any bifurcate horizon.  For asymptotically flat spaces, this ambiguity is removed by requiring that $k$ be normalised such that $k^2 \rightarrow -1$ as one approaches future null infinity.

Here, in defining the asymptotic symmetry group, we are using Gaussian null coordinates defined in the neighbourhood of the horizon.  Thus, in choosing the aforementioned coordinates and writing the metric in the form \eqref{NHC}, we have exorcised regions of the space away from the horizon.  Of course, this has the advantage that our analysis applies to any horizon, even if the original black hole is not asymptotically flat.  However, it means that we cannot use the above method to remove the ambiguity in the definition of the surface gravity.

Nevertheless, we proceed by defining the surface gravity associated with the horizon, written in Gaussian null coordinates to be that given by equation \eqref{kappa} with
\begin{equation}
 k = \partial_v.
\end{equation}
The reason for this definition is simply that any other normalisation of $k$ would be arbitrary.  In other words, we choose to remove the ambiguity using the very choice of Gaussian null coordinates, which distinguishes $v,$ as opposed to any other constant multiple of it.

Using the fact that
\begin{equation}
 k^2 = - L^2 r^n F, \qquad k^{\flat} = L^2 (dr - r^n F\, dv )
\end{equation}
and the fact that in these coordinates the horizon $\mathcal{H} = \{r=0\}$, equation \eqref{kappa} gives that
\begin{equation}
 \kappa = \begin{cases}
           \frac{1}{2} & n=1 \\
           0 & n=2
          \end{cases}.
\end{equation}
As expected the surface gravity vanishes for extremal horizons.  It may be argued that there is no significance in defining a surface gravity if it is going to be the same for all subextremal horizons.  However, as emphasised before, rather to the contrary, the significance of the surface gravity as defined above is tied to the definition of the Gaussian null coordinates in the vicinity of the horizon.  In other words, a distinguishing feature of Gaussian null coordinates may be thought of as those coordinates in which the surface gravity associated with a subextremal horizon is $\kappa = 1/2.$
In conclusion, we identify the exponential factors $e^{\pm v/2}$ that appear in the asymptotic symmetry generators for subextremal horizons, see equations \eqref{1:xiv}--\eqref{1:xir}, as redshift factors
\begin{equation}
 e^{\pm \kappa v}.
\end{equation}

\section{Example: spherically symmetric case} \label{sec:sph}

In section \ref{sec:ASG}, we have defined an asymptotic symmetry group associated for any Killing horizon.  However, it is important to note that given how we have defined the generators, there will be a unique symmetry group associated with any class of solutions depending on the choice of
\begin{equation*}
 L, \quad \tilde{h}^I, \quad \omega_{IJ}.
\end{equation*}

As an example we take the simple case where we have spherical symmetry.  Thus, we choose
\begin{equation}
 \tilde{h}^J = 0, \qquad  \gamma_{IJ} = \rho (r)\, \omega_{IJ},
\end{equation}
where now $\omega_{IJ}$ is the metric on the round 2-sphere and
\begin{equation}
 \rho(r) = 1 + \rho^{(1)} r + \mathcal{O}(r^2).
\end{equation}
Moreover, we choose $L$ to be constant.

This class of horizons includes the Schwarzschild and (extreme) Reissner-Nordstr\"om horizons.

As before we consider each case separately.  In this section, for brevity, we unambiguously raise and lower $I,J,\ldots$ indices using $\omega_{IJ}$ and its inverse.

\subsection*{n=2: Extremal case}

Equations \eqref{2:xirdef}, \eqref{2:Idef} and \eqref{2:beta} give that
\begin{equation}
  \hat{\xi}^r = \hat{\xi}^{r}(x), \qquad 
 \hat{\xi}^I = \alpha^I(x) + \beta^I(x) \, v,
\end{equation}
where
\begin{equation}
 \beta^I = - L^2 \omega^{IJ} \partial_{J} \hat{\xi}^r.
\end{equation}
Moreover, equation \eqref{2:betaKilling} implies that
\begin{equation}
 \nabla_{I} \nabla_{J} \hat{\xi}^r = 0,
\end{equation}
where, as before, $\nabla_I$ denotes a covariant derivative associated with the round 2-sphere metric. Imposing regular boundary conditions, this implies that
\begin{equation}
  \hat{\xi}^r = \textup{const.}
\end{equation}
and so $\beta^I=0.$
Now, equation \eqref{2:xiga1} implies that
\begin{equation}
 \hat{\xi}^r = 0.
\end{equation}
Then, equations \eqref{2:vdef} and \eqref{2:h} imply that
\begin{equation}
 \xi^v = f(x) + g(x) v,
\end{equation}
Equations \eqref{2:xir1} and \eqref{2:xir2} become trivial, while equation \eqref{2:xiga2} gives that $g=c,$ a constant.  Finally, equation \eqref{2:aKilling} reduces to the fact that $\alpha^I \equiv Y^I,$ a Killing vector field on the round 2-sphere.

In summary, 
\begin{gather}
 \xi = \big( f(x) + c\, v \big) \partial_v + \left\{ Y^I(x) - L^2 \omega^{IJ} \partial_{J}f \int dr\, \rho(r)^{-1} \right\} \partial_I - \left\{ c \, r + \partial_I f \int dr\ r h^I    \right\} \partial_r.
\end{gather}
Thus, the asymptotic symmetry group is generated by a constant $c$, an arbitrary function $f(x),$ the supertranslations and the three Killing vectors $Y^I$ on the round 2-sphere, which generate $SO(3).$

\subsection*{n=1: Subextremal case}

From equations \eqref{1:xirdef}, \eqref{1:Idef}, we read off
\begin{equation}
 \hat{\xi}^r = \zeta(x)\, e^{v/2}, \qquad
 \hat{\xi}^I = Y^I(x) - 2\,  e^{v/2}  L^2 \omega^{IJ} \partial_{J} \zeta.
\end{equation}
Equation \eqref{1:zeta1} reduces to
\begin{equation}
 \nabla_I \nabla_J \zeta = \frac{\rho^{(1)}}{4 L^2} \ \zeta.
\end{equation}
Thus, $\zeta$ is a spherical harmonic.  Imposing regular boundary conditions would imply that
\begin{equation}
 \rho^{(1)}/L^2 = -2 \ell (\ell + 1)
\end{equation}
for non-negative integer $\ell.$  Since, the left hand side is generic, the above equation cannot be satisfied for a generic space in this class.  Therefore, in order not to contradict the assumption that the space given by $\omega_{IJ}$ is compact, we must conclude that
\begin{equation}
 \zeta = 0.
\end{equation}
Now, equations \eqref{1:vdef} and \eqref{1:h} reduce to 
\begin{equation}
 \xi^v = f(x) e^{-v/2} + g(x).
\end{equation}
Equations \eqref{1:zeta1} and \eqref{1:zeta2} become trivial, while equation \eqref{1:gY} implies that $g=c,$ a constant.  Moreover, from equation \eqref{1:Y}, we have that $Y^I$ is a Killing vector field on the round 2-sphere.

In summary, 
\begin{gather}
\xi = c\, \partial_v +  Y^I(x) \partial_I \hspace{110mm} \notag \\[2mm]
 \hspace{10mm} + e^{-\kappa v} \left\{  f(x) \, \partial_v - \left[L^2  \omega^{IJ} \partial_{J}f \int dr\, \rho(r)^{-1} \right] \partial_I 
 +  \left[ \frac{1}{2} f\,  r - \partial_I f \int dr\ r h^I    \right]  \partial_r \right\}.
\end{gather}
As before, for the extremal case, the asymptotic symmetry group is generated by a constant $c$, an arbitrary function $f(x),$ the supertranslations and the three Killing vectors $Y^I$ on the round 2-sphere.  However, as emphasised before, there is a significant difference in the $v$-dependence of the generators.  As was discussed in section \ref{sec:kappa}, this is related to the fact that observers experience a gravitational redshift effect for subextremal horizons, whereas extremal horizons admit no such property.

\section{Discussions} \label{sec:dis}

We have determined the asymptotic symmetry group of Killing horizons in four dimensions.  We take as the background space, about which the symmetry analysis is performed, the space given by the leading order terms in the metric in Gaussian null coordinates near the horizon.  This is analogous to the asymptotically flat case, where Bondi coordinates are chosen and the background about which the asymptotic symmetry analysis is performed is flat Minkowski space.  As is the case with the asymptotic symmetry group there, the BMS group, here also, we have supertranslations along the horizon.

An important step in terms of the physical understanding of the asymptotic symmetry generators we find here and the relation to the quantum theory is a Hamiltonian analysis of the symmetry group and the corresponding charges (see e.g.\ \cite{Barnich:2011mi, Compere:2011ve, Troessaert:2017jcm}).  We hope to return to this issue in the near future.

Another interesting question that arises and would be naturally related to the Hamiltonian analysis discussed above is the relation of the asymptotic symmetry group for extremal horizons and the Aretakis charges \cite{Aretakis:extremal}. Is it clear from the structure of the corresponding symmetry groups why Aretakis charges should exist for extremal, and not for subextremal, horizons?  Even for the case of Newman-Penrose charges \cite{NP} at future null infinity, it is known that their existence is not inextricably linked to the existence of BMS symmetry \cite{GGP}.  Therefore, the relation between the asymptotic symmetry charges found here and the Aretakis charges are not expected to be direct.   

\section*{Acknowledgements}
We would like to thank Hadi Godazgar and Malcolm Perry for useful discussions.  We would like to thank the Max-Planck-Institut f\"ur Gravitationsphysik (Albert-Einstein-Insitut), Potsdam, for kind and generous hospitality during the course of this work. E.T.A.\ is partially supported by the Russian state grant Goszadanie 3.9904.2017/BCh.  M.G.\ is partially supported by grant no. 615203 from the European Research Council under the FP7.

\bibliographystyle{utphys}
\bibliography{asymp}

\end{document}